\documentclass[aps,prl,preprint,showpacs,preprintnumbers,amsmath,amssymb]{revtex4-1}
\usepackage{graphicx}
\usepackage{dcolumn}
\usepackage{bm}

\usepackage{multirow}
\usepackage{epsfig}
\usepackage{amssymb}

\begin{document}
\title{Structural Evolution and Optoelectronic Applications of Multilayer Silicene}

\author{Zhi-Xin Guo$^{1}$$^{\ast }$, Yue-Yu Zhang$^{2}$, Hongjun Xiang$^{2}$, Xin-Gao Gong$^{2}$ and Atsushi Oshiyama$^{3}$}

\affiliation{$^{1}$ Department of Physics, Xiangtan University, Xiangtan, Hunan 411105, China\\
$^{2}$ Key Laboratory of Computational Physical Sciences (Ministry of Education), State Key Laboratory of Surface Physics, Collaborative Innovation Center of Advanced Microstructures, and Department of Physics, Fudan University, Shanghai 200433, P. R. China\\
$^{3}$ Department of Applied Physics, The University of Tokyo, Hongo, Tokyo 113-8656, Japan}
\email{zxguo08@hotmail.com}



\date{\today}

\begin{abstract}
Despite the recent progress on two-dimensional multilayer materials (2DMM) with weak interlayer interactions, the investigation on 2DMM with strong interlayer interactions is far from its sufficiency. Here we report on first-principles calculations that clarify the structural evolution and optoelectronic properties of such a 2DMM, multilayer silicene. With our newly developed global optimization algorithm, we discover the existence of rich dynamically stable multilayer silicene phases, the stability of which is closely related to the extent of $sp^{3}$ hybridization that can be evaluated by the average bonds and effective bond angles. The stable Si(111) surface structures are obtained when the silicene thickness gets up to four, showing the critical thickness for the structural evolution. We also find that the multilayer silicene with $\pi$-bonded surfaces present outstanding optoelectronic properties for the solar cells and optical fiber communications due to the incorporation of $sp^{2}$-type bonds in the $sp^{3}$-type bonds dominated system. This study is helpful to complete the picture of structure and related property evolution of 2DMM with strong interlayer interactions.
\end{abstract}


\maketitle
Two-dimensional multilayer material (2DMM) is a kind of important nanomaterial. By far 2DMM with weak interlayer interactions, such as graphene, MoS$_{2}$, black phosphorus, etc., have been extensively investigated for their fascinating properties\cite{geim,appenzeller,zhangy}, whereas those with strong interlayer interactions are barely explored. The two-dimensional structure of Si (silicene), which has been both theoretically predicted and experimentally synthesized\cite{cahangirov,vogt,lin,fleurence,feng,chen,gao,padova,guo1,guo2,jia}, can be an ideal material for such investigation.
Despite the extensive studies on monolayer silicene\cite{yao,ezawa,ni,pan,tao,oughaddou}, the investigation on multilayer silicene is far from its sufficiency.
Although a few theoretical investigations have been performed on the freestanding bilayer silicene\cite{zeng,meng,xiang,sakai}, the mechanism for the relative stability among the obtained phases has not been understood.  Moreover, bilayer silicene can be either metal or semiconductor depending on the specific stacking structures\cite{meng,xiang,sakai}, while the underlying mechanisms have not been clarified.
Also, the structural and electronic properties of thicker freestanding multilayer silicene are still to be explored.

On the other hand, Si surfaces are premier stages on which the fabrication of most electron devices is achieved.
The atomic layers of Si(111) surface has the hexagonal network in the lateral plane, and thus the similarity and dissimilarity between the multilayer silicene and the surface atomic layers of Si(111) surface are intriguing. Actually the cleaved Si(111) surface shows $2 \times 1$ periodicity and then its annealing converts it to the more stable $ 7 \times 7$ periodicity. The $2 \times 1$ phase has been identified as the $\pi$-bonded chain structure \cite{pandey} and a dimer-adatom-stacking-fault (DAS) model \cite{takayanagi,takayanagi1} is now established for the $ 7 \times 7 $ phase \cite{stich,brommer}. It is important and interesting to clarify structural evolution between the silicene and reconstructed surface phases in the multilayer Si.

Although the majority of solar cells fabricated to date have been based on the three-dimensional (3D) Si, it is well known that 3D Si is not an ideal material for optoelectronic applications due to its large direct band gap (3.4 eV\cite{hybertsen}). Because of the excellent compatibility with the mature Si-based microelectronics industry and high abundance of Si, the discovery of multilayer silicene materials with excellent optoelectronic properties may lead to revolution in future optoelectronics technology.

In this work, by performing extensive calculations for the multilayer silicene from bilayer to quadlayer, we clarify the structural evolution and the outstanding optoelectronic properties of multilayer silicene. Searches for the stable multilayer silicene phases are performed using our newly developed global optimization algorithm code (IM$^{2}$ODE\cite{im2}), which is developed based on Differential Evolution. We consider the structures with a different number $n$ of Si atoms in the unit cell ($2\leq n\leq36$). The population size is set to be 20 and the number of generations ia fixed at 15 in usual cases. The structural relaxations are performed using the Vienna \textit{ab initio} simulation package (VASP)\cite{vasp1}. The stability of the obtained phases are verified by the phonon calculations\cite{phonon}. The final accurate calculations of the energy bands and optical spectra are performed using the HSE06 functional as implemented in VASP.

\begin{figure}
\begin{center}
\includegraphics[angle= 0, width=0.98\linewidth]{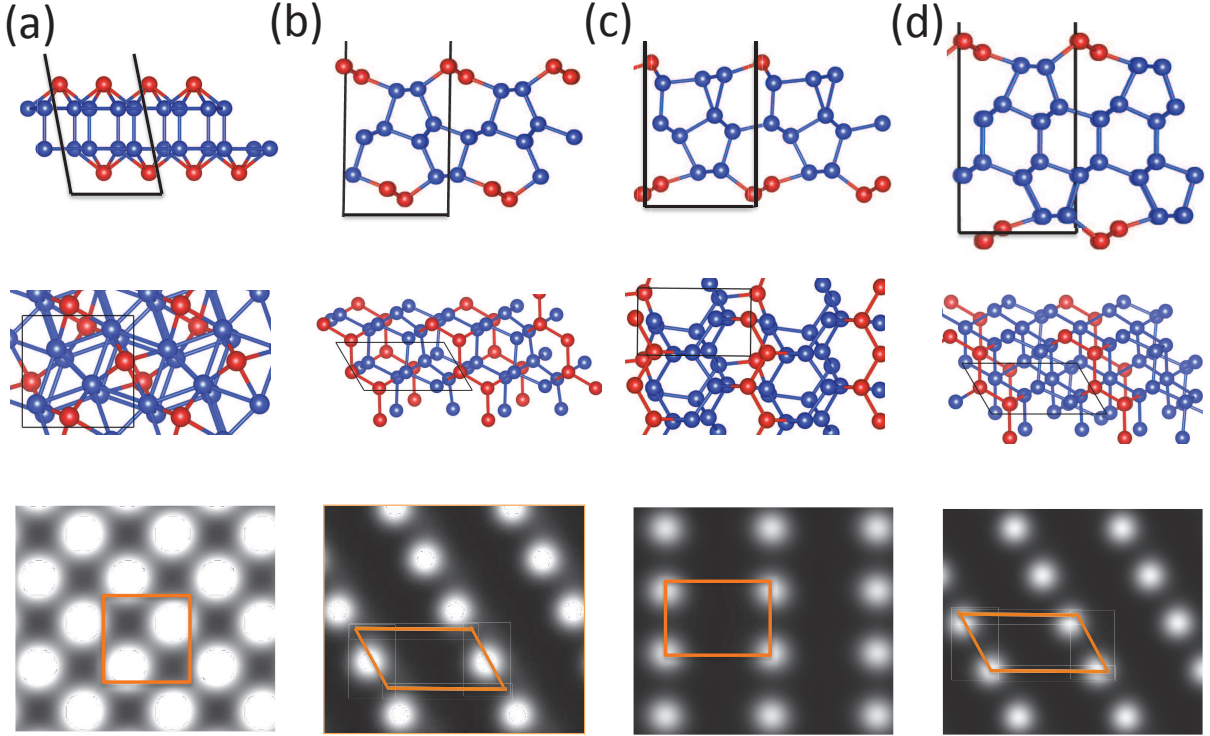}
\caption{
Side (upper panels) and top (middle panels) views of stable phases of freestanding multilayer silicene and corresponding occupied STM images (bottom panels). (a) C12/m1-$\sqrt{2}\times\sqrt{2}$ bilayer silicene, (b) P121/m1-$2\times1$ and (c) P1-$2\times1$ trilayer silicene, (d) P1m1-$2\times1$ quadlayer silicene.
The highly protruded Si atoms at the surfaces are depicted by the red balls and the remaining Si atoms are depicted by the blue balls. The lateral unit cells are also indicated by the solid lines.
}
\label{str}
\end{center}
\end{figure}

With the IM$^{2}$ODE code, we have performed extensive searches for the most stable phases of silicene from monolayer to quadlayer. We have obtained several new monolayer silicene phases [Fig. S1] which are energetically stable to the well-known low-buckled honeycomb phase\cite{cahangirov}, but they are unstable from the phonon calculations. This shows that the low-buckled honeycomb structure is the most stable monolayer silicene phase.

The most stable bilayer silicene phase we have obtained is the hex-OR-$2\times2$ structure with the cohesive energy ($E_{c}$) being 5.073 eV/Si [Fig. S2(a)]\cite{SM}, which is the same as that obtained in previous work using the structural exploration method\cite{sakai}. Since this phase presents the P-1 symmetry, here we denote it as the P-1-$2\times2$ phase. We have also reached the slide-$2\times2$ ($E_{c}$=5.063 eV/Si) and Cmme-$1\times1$ ($E_{c}$=5.000 eV/Si) phases as previously reported\cite{xiang,sakai}, which are the secondary most stable and the most stable phases with periodicities of $2\times2$ and $1\times1$, respectively [Fig. S2(b) and S2(c)]\cite{SM}. The slide-$2\times2$ phase has the C12/m1 symmetry, thus it is denoted as C12/m1-$2\times2$ phase accordingly.
Additionally, we have obtained a new phase which presents the C12/m1 symmetry with periodicity of $\sqrt{2}\times\sqrt{2}$ ($E_{c}$=4.991 eV/Si, denoted as C12/m1-$\sqrt{2}\times\sqrt{2}$) [Fig. \ref{str}(a)]. The four bilayer silicene phases are dynamically stable as evidenced from the calculated phonon spectra shown in Fig. S3(a)\cite{SM} as well as in Ref.\onlinecite{sakai}. The cohesive energies of the bilayer silicene phases are larger than the monolayer silicene by 240-320 meV/Si, showing the covalent interlayer interactions.
The above results show the existence of rich stable bilayer silicene phases. The calculated occupied STM\cite{stm} images of the four bilayer silicene phases are different from each other with different periodicities, i.e., the hexagonal, tetragonal, and rhombic geometries for the $1\times1$, $\sqrt{2}\times\sqrt{2}$, and $2\times2$ phases respectively, which make them easily distinguished in experiments. Whereas, the two rhombic STM images of the P-1-$2\times2$ and C12/m1-$2\times2$ phases are very similar, additional techniques have to be used to distinguish them.

To explore the structural evolution in multilayer silicene, we have also performed extensive searches for the stable trilayer silicene phases. Several stable phases have been obtained, where the most stable two present a $2\times1$ periodicity with the hexagonal ($E_{c}$=5.138 eV/Si) and rectangular supercell ($E_{c}$=5.135 eV/Si) [Figs. \ref{str}(b) and \ref{str}(c)], respectively. The dynamical stability of them has also been verified by the phonon calculations [Figs. S3(b) and S3(c)]\cite{SM}. The hexagonal and rectangular phases have the P121/m1 and P1 symmetries, they are thus denoted as P121/m1-$2\times1$ and P1-$2\times1$, respectively.
Particularly, in the P121/m1-$2\times1$ phase the top (bottom) silicene layer is drastically reconstructed to form chains of $\pi$ orbitals associated with the five- and seven-membered rings of the top (bottom) and the middle silicene layers. This feature is the same as that of $2\times1$ Si(111) surface, but there is obvious distortion in the middle silicene layer with respect to the second Si layer\cite{silayer} of the $2\times1$ Si(111) surface \cite{pandey}. The structure of P1-$2\times1$ phase is similar with the P121/m1-$2\times1$ phase. The difference mainly lies at that the seven-membered ring in the P121/m1-$2\times1$ phase splits into the three-membered and six-membered rings in the P1-$2\times1$ phase. The calculated occupied STM images of both phases are shown in Figs. \ref{str}(b) and \ref{str}(c), which present the atomic chain image as that of the $2\times1$ Si(111) surface.
We have also verified the stability of the trilayer $7\times7$ silicene phase with the initial surface structure being the same as the $7\times7$ Si(111) surface with DAS model\cite{takayanagi,takayanagi1}. After geometry optimization, however, the geometry distorts a lot. This shows that trilayer silicene is too thin to reproduce the complex $7\times7$ Si(111) surface structure.

We have then explored the stable phases for the thicker silicene, i.e., the quadlayer silicene. With the IM$^{2}$ODE code, we have obtained several stable phases, among which the $2\times1$ phase with P1m1 symmetry is the most stable with $E_{c}$=5.225 eV/Si [denoted as P1m1-$2\times1$, Fig. \ref{str}(d)]. The top and bottom two silicene layers in this phase present the perfect $\pi$-bonded chain structure, the geometries of which resemble to the $2\times1$ Si(111) surfaces with positive and negative buckling reconstructions \cite{zitzlsperger}, respectively. The dynamical stability of the $2\times1$ silicene phase is verified by the phonon calculations [Fig. S3(d)]\cite{SM}. We have also verified the stability of the $7\times7$ DAS structure following the way in the trilayer silicene. After geometry optimization, the stable $7\times7$ DAS structure is obtained in the quadlayer silicene [Fig. S2(d)].
The simulated occupied STM images of the $2\times1$ [Fig. 1(d)] and $7\times7$ [Fig. S2(d)]\cite{SM} silicene phases confirm their structural similarity to those of the Si(111) surfaces.
The $E_{c}$ of stable $7\times7$ phase is 5.245 eV/Si, 20 meV/Si larger than the $2\times1$ phase. This is consistent with the fact that the $7\times7$ reconstruction is more stable than the $2\times1$ reconstruction in Si(111).
The above results show that quadlayer is the critical thickness for the structural transition from silicene to Si(111) surface.

Although multilayer silicene phases are rich and complex, their structures can be roughly characterized by using the average bond number (ABN)\cite{ABN} and average effective bond angle (AEBA)\cite{AEBA}. For instance, the ABN in both Cmme-$1\times1$ and C12/m1-$\sqrt{2}\times\sqrt{2}$ bilayer silicene is 4.00 since all the Si atoms are fourfold-coordinated, whereas it is 3.75 in the P-1-$2\times2$ and C12/m1-$2\times2$ bilayer silicene since 25\% Si atoms are threefold-coordinated and the remaining 75\% Si atoms are fourfold-coordinated (Table. 1).
The AEBA can be normalized with respect to the ideal $sp^{3}$ (AEBA1) and $sp^{2}$ (AEBA2) bond angles which are $109.47^{o}$ and $120^{o}$, respectively\cite{AEBA}. From Table. 1, for all the multilayer silicene phases, the ABN is larger than 3.50 and the AEBA1 is larger than AEBA2. This feature shows that the majority of Si atoms are fourfold-coordinated and the bond angles are more close to the $sp^{3}$ type, reflecting the nature of $sp^{3}$-like hybridization in multilayer silicene. It is noteworthy that the ABN equals to 3.00 and the AEBA1 is smaller than AEBA2 in monolayer silicene, showing the nature of $sp^{2}$-like hybridization.

The cohesive energy $E_{c}$ is summarized in Table. 1 to show the stability of the silicene phases.
A common feature is that $E_{c}$ increases with the silicene thickness increasing, which is 240-480 meV/Si stable to the monolayer silicene. This shows that thicker silicene is easier obtained in experiments.
It is noticed that the P-1-$2\times2$ and C12/m1-$2\times2$ bilayer silicene are more stable than the Cmme-$1\times1$ and C12/m1-$\sqrt{2}\times\sqrt{2}$ bilayer silicene by 60-70 meV/Si, which is unexpected since the ABN is smaller by 0.25 bond per Si in the former two phases. This is mainly own to the different bond angle distributions, which affect the strength of $s$-$p$ hybridizations and thus the bond energies. The AEBA1 is larger in the two $2\times2$ phases than those in the $1\times1$ and $\sqrt{2}\times\sqrt{2}$ phases, showing that the structural characteristics in the $2\times2$ phases are more close to the diamond Si which is the most stable Si phase. The similar tendency is also obtained in thicker silicene phases (Table. 1). These results imply a general rule for the stability of dynamically stable multilayer silicene with similar ABN values, i.e., the silicene phases with larger AEBA1 values are usually more stable.

\begin{table}
\caption{
Geometry symmetry, cohesive energy $E_{\rm c}$ (eV/Si), average bond number (ABN), average effective bond angle normalized with respect to $109.47^{o}$ (AEBA1) and $120^{o}$ (AEBA2), as well as direct band gap (DBG) in unit of eV for monolayer (ML), bilayer (BL), trilayer (TL) and quadlayer (QL) silicene.}
\label{structure1}
\begin{tabular}{cccccccc}
  \hline
  \hline
&Periodicity  &Symmetry   & $E_{c}$  & ABN   &AEBA1 & AEBA2    &DBG\\
  \hline
  ML         &$1\times1$      & P6/mm         & 4.750        & 3.00       &0.94        &0.96      &--  \\
 BL           &$1\times1$       & Cmme         & 5.000        & 4.00        &0.87        &0.83     &--  \\
 BL           &$\sqrt{2}\times\sqrt{2}$    & C12/m1       & 4.991        & 4.00       &0.85        &0.82    &-- \\
 BL           &$2\times2$       & P-1          & 5.073        & 3.75       &0.89        &0.86    &1.5  \\
 BL         &$2\times2$       & C12/m1       & 5.063        & 3.75        &0.88        &0.86     &0.5  \\
 TL        &$2\times1$       &P121/m1      &5.138         & 3.67        &0.93        &0.90     &0.5 \\
 TL          &$2\times1$       &P1           &5.135          & 3.83        &0.91        &0.88     &0.9   \\

  QL          &$2\times1$       &P1m1      & 5.225        & 3.75         &0.94        &0.90     &0.7    \\
  \hline
  \hline
\end{tabular}
\end{table}

\begin{figure}
\begin{center}
\includegraphics[angle= 0,width=0.98\linewidth]{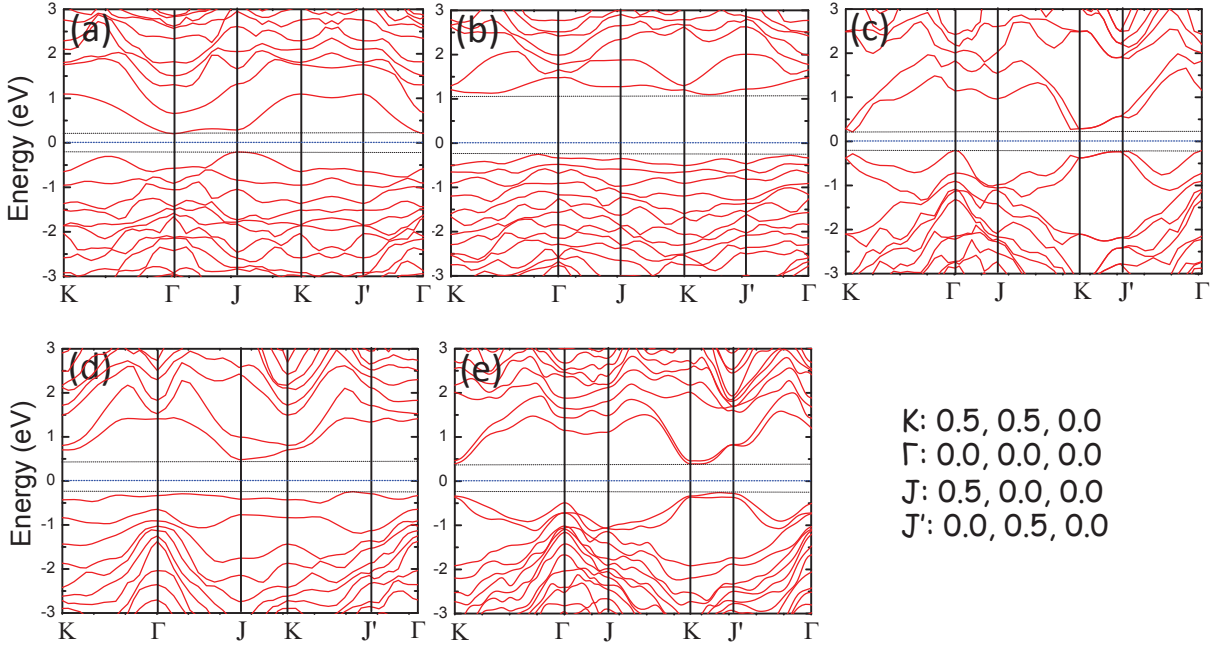}
\caption{
Calculated energy bands of semiconductor multilayer silicene. (a) Energy bands of C12/m1-$2\times2$ bilayer silicene, (b) Energy bands of P-1-$2\times2$ bilayer silicene, (c) Energy bands of P121/m1-$2\times1$ trilayer silicene, (d) Energy bands of P1-$2\times1$ trilayer silicene, (e) Energy bands of P1m1-$2\times1$ quadlayer silicene. The valence band maximum (VBM) and conduction band minimum (CBM) are indicated by the short-dotted black lines.
}
\label{bd}
\end{center}
\end{figure}

The electronic properties of the multilayer silicene phases are further calculated. It is found that the Cmme-$1\times1$ and C12/m1-$\sqrt{2}\times\sqrt{2}$ bilayer silicene are metallic (Fig. S4)\cite{SM}, whereas the P-1-$2\times2$ and C12/m1-$2\times2$ bilayer, the P121/m1-$2\times1$ and P1-$2\times1$ trilayer, as well as the P1m1-$2\times1$ quadlayer silicene are indirect semiconductors with the band gaps varying from 0.4 to 1.3 eV (Fig. \ref{bd}).
From the structural details, the $s$-$p$ orbital hybridizations in the former two metallic phases are distorted a lot from the ideal $sp^{3}$ hybridization although all the Si atoms are fourfold-coordinated. As for the later four semiconductor phases, however, the hybridizations of threefold-coordinated and fourfold-coordinated Si atoms resemble to the $sp^{2}$ and $sp^{3}$ hybridizations, respectively. The different form of orbital hybridizations in the multilayer silicene induce the very different electronic properties.
These results indicate that the electronic properties of multilayer silicene can be effectively manipulated by the structural engineering.
On the other hand, the direct band gaps (DBG) of these indirect semiconductor silicene phases are in range of 0.5-1.5 eV (Table. 1), being very close to their indirect band gaps. This implies the outstanding optoelectronic properties of them for the applications in the optical fiber communications and the solar cells, where the required optimal band gaps are 0.8 and 1.5 eV, respectively.

It is well known that the imaginary part of the dielectric tensor $\varepsilon_{2}$ determines the absorption ability of a material. The calculated $\varepsilon_{2}$ of the semiconductor multilayer silicene phases are shown in Fig. S5\cite{SM}. After careful analyses, it is found that the optical absorptions in all these phases start at the direct band gap. This implies that the P-1-$2\times2$ bilayer silicene (DBG=1.5 eV) and P1-$2\times1$ trilayer silicene (DBG=0.9 eV) are suitable for the solar cells according to the Shockley-Queisser limit\cite{shock}, whereas the C12/m1-$2\times2$ bilayer silicene (DBG=0.5 eV), P121/m1-$2\times1$ trilayer silicene (DBG=0.5 eV) and P1m1-$2\times1$ quadlayer silicene (DBG=0.7 eV) are suitable for the fiber communications.

\begin{figure}
\begin{center}
\includegraphics[angle= 0,width=0.98\linewidth]{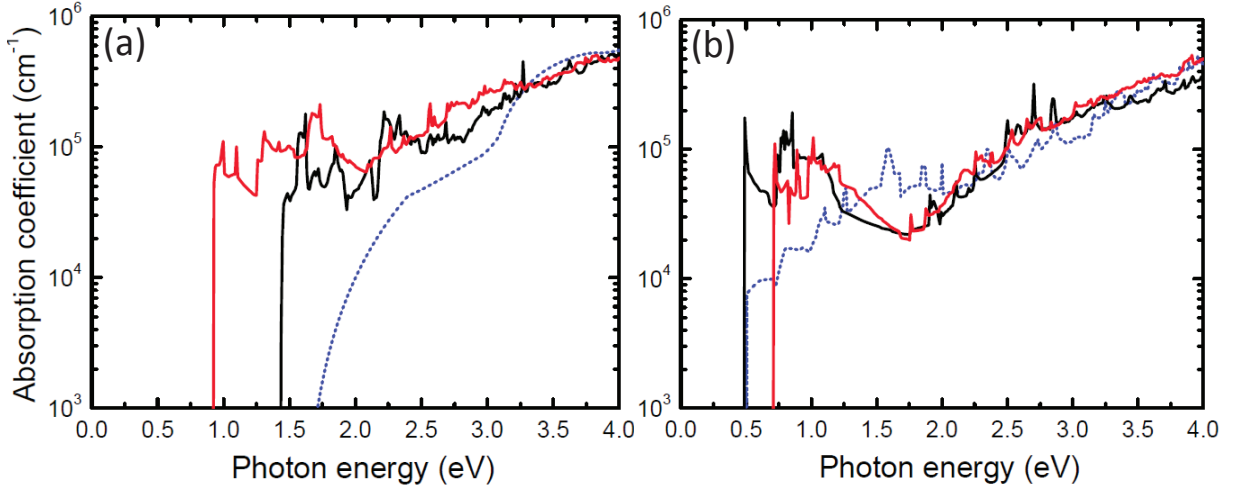}
\caption{
Calculated optical absorption coefficients of multilayer silicene with the compare of GaAs. (a) Optical absorption coefficients of P-1-$2\times2$ bilayer silicene (solid black line), P1-$2\times1$ trilayer silicene (solid red line), and bulk GaAs (short-dotted blue line), (b) Optical absorption coefficients of C12/m1-$2\times2$ bilayer silicene (short-dotted blue line), P121/m1-$2\times1$ trilayer silicene (solid black line), and P1m1-$2\times1$ quadlayer silicene (solid red line).
}
\label{absorb}
\end{center}
\end{figure}

We have further calculated the optical absorption coefficients of the semiconductor multilayer silicene phases and compared them with that of GaAs, the conversion efficiency of which is the highest among all the thin-film solar cell absorbers. As shown in Fig. \ref{absorb}(a), the optical absorption coefficients of P-1-$2\times2$ bilayer silicene and P1-$2\times1$ trilayer silicene are up to $\sim 10^{5} cm^{-1}$, one order of magnitude higher than that of GaAs in the energy range of about 1.0-2.0 eV, showing that they are very good nanoscale solar cell absorbers. On the other hand, the absorption coefficients of P121/m1-$2\times1$ trilayer silicene and P1m1-$2\times1$ quadlayer silicene are also up to $\sim 10^{5} cm^{-1}$ in energy range of 0.5-1.0 eV [Fig. \ref{absorb}(b)], showing that they are good nanoscale fiber communications. While the absorption coefficient of the C12/m1-$2\times2$ bilayer silicene is one order smaller [Fig. \ref{absorb}(b)], indicating that it is not suitable for the optical applications.

\begin{figure}
\begin{center}
\includegraphics[angle= 0,width=0.98\linewidth]{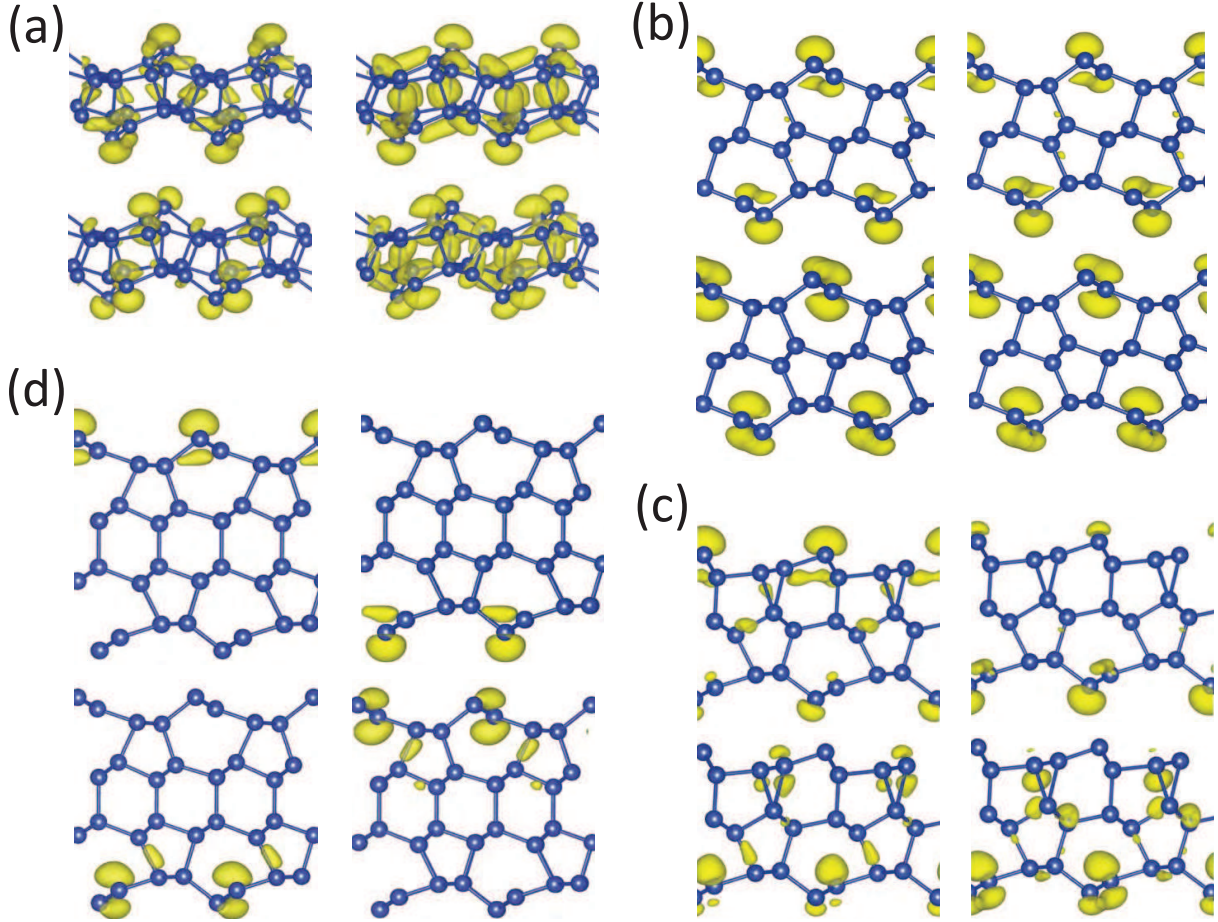}
\caption{
Squared Kohn-Sham (KS) orbitals of the first two valence (upper panels) and conduction (bottom panels) bands shown in Fig. \ref{bd} at $K$ point in Brillouin zone for (a) P-1-$2\times2$ bilayer silicene [Fig. \ref{bd}(b)], (b) P121/m1-$2\times1$ trilayer silicene [Fig. \ref{bd}(c)], (c) P1-$2\times1$ trilayer silicene [Fig. \ref{bd}(d)], and (d) P1m1-$2\times1$ quadlayer silicene [Fig. \ref{bd}(e)]. The isovalue surface of the squared KS orbital is taken as 30\% of the maximum value.
}
\label{wf}
\end{center}
\end{figure}

To explore the mechanism of the outstanding optoelectronic properties of multilayer silicene, we have analyzed the corresponding optical transition matrix and the energy bands of them. It is found that the absorption peeks with photon energy smaller than 2.0 (1.0) eV for P-1-$2\times2$ bilayer silicene and P1-$2\times1$ trilayer silicene (P121/m1-$2\times1$ trilayer silicene and P1m1-$2\times1$ quadlayer silicene) are mainly attributed to the transitions between the first two valence and conduction bands, especially along the $K-J'$ direction in the Brillouin zone. The Kohn-Sham (KS) orbitals (Fig. \ref{wf}) further show that these energy bands mainly come from the $p_{z}$ ($\pi$) orbitals of the Si atoms at silicene surfaces, which present $sp^{2}$ hybridization. Similar results have also been found in the thicker silicene phases, i.e., the 6- and 8-Si layered $2\times1$ silicene (Fig. S6\cite{SM}).
These results clearly show that the promising optical properties of these multilayer silicene phases are induced by the incorporation of $sp^{2}$ hybridization. It is known that the materials consisted of pure $sp^{2}$-type (such as monolayer graphene, silicene) and $sp^{3}$-type (such as diamond C, Si) bonds with group IV elements usually have too small and too large direct band gaps for the optoelectronic applications, respectively.
A common feature for the multilayer silicene phases with appropriate direct band gaps is the existence of some $sp^{2}$-type bonds in the $sp^{3}$-type bonds dominated system.
This kind of configuration can therefore overcome the deficiency on the energy bands of a monotonous $s$-$p$ hybridized system.

In conclusion, by combining our newly developed global optimization algorithm with first-principles calculations, we have discovered many dynamically stable multilayer silicene phases with different thicknesses and explored the mechanism for their relative stability. We have clarified that quadlayer is the critical thickness for the structural evolution from silicene to Si(111). Finally, we have found that the multilayer silicene with $\pi$-bonded surfaces present outstanding optoelectronic properties due to the incorporation of $sp^{2}$-type bonds in the $sp^{3}$-type bonds dominated system, which lightens an efficient way to the material design for the optoelectronic devices. These findings are also helpful to complete the picture of structure and related property evolution for the strong interlayer-interacted 2DMM.

\begin{acknowledgments}
We acknowledge Mr. Wei Luo for helpful discussions. Work at Xiangtan was supported by NSFC (Grant No. 11204259), the NSF of Hunan Province (Grant No. 2015JJ6106); Work at Fudan was supported by NSFC, Special Funds for Major State Basic Research; Work at Tokyo was supported by the research project ``Materials Design through Computics" (http://computics-material.jp/index-e.html) by MEXT and the ``Computational Materials Science Initiative" by MEXT, Japan.
\end{acknowledgments}


\begin{thebibliography}{}
\bibitem{vasp1}
The exchange-correlation functional, vdW-DF (optB86b-vdW functional)\cite{dion,klimes}, being capable of treating the dispersion force is adopted. The electron-ion interaction is described by the projector augmented wave method\cite{paw}, and the cutoff energy of 250 eV in the plane-wave basis set is used. The Monkhorst-Pack grid with a maximum spacing of $2\pi\times0.02$ \AA$^{-1}$ is employed in the calculations. All atomic positions and lattice constants are optimized by using the conjugate gradient method where total energy and atomic forces are minimized. The geometry optimization is performed until the remaining Hellmann-Feynman forces become less than 0.001 eV/{\AA} to obtain the final structures.
\bibitem{phonon}
Phonon calculations are performed using the force-constant method as implemented in phononpy program with forces calculated by VASP.
\bibitem{silayer}
A Si layer is defined following the convention of the silicene layer, i.e., a single bilayer layer, as consisting of 2 Si atoms at the upper positions and the remaining 2 Si atoms at the lower positions in the $2\times 1$ lateral cell of the Si(111) surface.
\bibitem{ABN}
ABN is definded as $N_{bond}/n$, where $N_{bond}$ and $n$ are the total number of Si-Si chemical bonds (with Si-Si distance being smaller than 2.7 \AA) and Si atoms in silicene, respectively.
\bibitem{AEBA}
AEBA is defined as $1-[\underset{i=1}{\overset{N_{angle}}\sum}\left|{A-Ang(i)}\right|] / (A \cdot N_{angle})$, where $N_{angle}$ and $Ang(i)$ are the total number of bond angles and the value of $i_{th}$ bond angle in silicene, respectively. AEBA1 and AEBA2 correspond to A=$109.47^{o}$ and A=$120^{o}$, respectively.
\end{thebibliography}
\end{document}